\documentclass[letterpaper,twocolumn,prb,showpacs,floatfix,
superscriptaddress]{revtex4}
\usepackage{graphicx}
\usepackage{array}
\begin{document}
\title{A statistical mechanics approach to the factorization problem}

\author{P. Henelius}
\affiliation{Theoretical Physics, Royal Institute of Technology,
SE-106 91 Stockholm, Sweden}

\author{S. Girvin}
\affiliation{Department of Physics, Yale University, P. O. Box 208120,
  New Haven, CT 06520-8120}

\date{\today}

\begin{abstract}
  We map the problem of finding the prime factors of an integer to the
  statistical physics problem of finding the ground state of a
  long-range Ising-like model. As in the strongly disordered
  Newman-Stein (NS) spin-glass model, the bond distribution is
  exponentially wide and grows with system size, but unlike the NS
  model we find that it is not wide enough for a greedy algorithm to
  be applicable. On the other hand, we also find that the frustration
  and exponential width of the bond distribution renders classical and
  quantum annealing and tempering methods no faster than a random
  search for this challenging model.
  \end{abstract}

\maketitle

\section{Introduction}
Since the security of public-key cryptography depends on the presumed
difficulty of computing the factors of large integers, there is very
strong incentive to find efficient factorization algorithms. Shor's
quantum algorithm\cite{shor97} which can find the factors in a time
that scales polynomially with the system size has motivated a massive
research effort into quantum computing. While there are no known  %SMG0202 added known
classical
algorithms that scale polynomially, the state-of-the-art general
number field sieve scales sub-exponentially.\cite{lens93, pome96} The
time required to factor a composite integer q scales as $\exp[(\log
q)^\frac{1}{3}(\log\log q)^\frac{2}{3}]$, and this has enabled
factoring of, for example, the RSA-200 (200 digits, 663 bits)
challenge in 2005.

Our aim in this paper is to view the problem from a statistical
physics point of view.  
We show that the problem of finding two prime
factors of a large composite integer can be transformed to an
optimization or statistical mechanics problem in which Ising spins are
used to represent the numbers in binary format.  The resulting
expression can be treated as a system of interacting two-state spins,
similar to an Ising model in statistical physics with frustrated
long-range multi-spin interactions.  The model can be formulated so
that the desired prime factors are given by the lowest energy
configuration.\cite{peng08,garc10}    %SMG0202  I think, to be fair, we need to cite them early.

For statistical mechanics problems without frustration and with
moderate or no disorder, a number of very efficient Monte Carlo
algorithms have been developed.  On the other hand, certain special
classes of disordered optimization problems, such as the minimum
spanning tree problem,\cite{krusk56, read05} are exactly solved by
simple "greedy" algorithms, where locally optimal choices lead to a
global solution.  Somewhat paradoxically, a broader class of problems
is approximately soluble by greedy algorithms in the limit of extreme
disorder.  For example, the resistance of a random resistor
network\cite{read05} can be solved to a good approximation by
selecting the single cheapest path that percolates across the cluster.
This is asymptotically exact in the limit of infinite
disorder. Finding the ground state of certain models of classical spin
glasses\cite{newm94, jack10} are also minimum spanning tree problems
in the limit of very broad disorder. For quantum spin problems,
asymptotically exact renormalization group methods based on analogous
ideas have been developed\cite{bhat82,fish94} Here the pair of spins
with the single strongest bond is integrated out which perturbatively
modifies the remaining bonds.  As long as the bond distribution
continues to broaden under repetition of this renormalization
procedure, the system flows to an infinite-disorder fixed point and
this treatment is asymptotically exact.

As we will see below, the factoring problem presents several extreme
difficulties.  First we require the exact ground state.  Nearby
solutions are of no use.  Second, the problem contains multi-spin
interactions and the bond strengths, while not truly random as in a
spin-glass model, are frustrated and very broadly distributed, but not
so broadly that a greedy algorithm is even approximately applicable.
Third, the energy landscape is extremely rough and there is no concept
of nearness in spin space.  Two nearly identical spin configurations
can have vastly different energies and conversely two very different
configurations can have nearly identical energies.

Our main result is the formulation of the problem in a spin language
which sheds light on why this is such a difficult optimization
problem. We explore a number of numerical methods for attacking this
model, including greedy algorithms, single and multi-spin updates,
parallel tempering and quantum annealing, However we find that all
exhibit exponential cost and, being essentially no faster than a
random search, are completely defeated by this difficult model.

\section{The Ising model}

In this section we discuss the form of the two-state model
representing the factorization problem. Assume that we want to find
two (unknown) prime factors $p_1$ and $p_2$ of a composite integer
$q$.  We can write the factors in binary form, $p_1=\sum_{i=0}^{n-1}
2^iS^1_i$ and $p_2=\sum_{j=0}^{n-1} 2^jS^2_j$, where the spin
variables $S\in\{0,1\}$. Considering two $n$-spin factors, the
factorization problem, $q=p_1p_2$, can be written
\begin{equation}
q=\sum_{i,j=0}^{n-1}2^{i+j}S^1_iS^2_j,
\label{factor}
\end{equation}
The right hand side is of the same form as a long-range Ising model,
$H=\sum_{i,j}J_{i,j}S^1_iS^2_j$, with the exponential coupling constant
$J_{i,j}=2^{i+j}$. Note that only spins belonging to separate integers
interact. We can now use Eq.~(\ref{factor}) to construct a cost
function that can be used as an effective Hamiltonian in a Monte Carlo
simulation. The ground state should be given by the desired prime
factors and first we consider
\begin{equation}
H_m=\left|q-\sum_{i,j=1}^n2^{i+j}S^1_iS^2_j\right|^m,
\label{h1}
\end{equation}
%\begin{equation}
%H_2=(q-\sum_{i,j=1}^n2^{i+j}S^1_iS^2_j)^2,
%\label{h2}
%\end{equation}
where $m$ is a positive number.  The ground state is doubly degenerate
due to a simple permutation of $p_1$ and $p_2$. We will consider the
cases $m=1/2,1,2$.  The case $m=2$ has the simplest interpretation as
a statistical mechanics problem since it corresponds to a model with
two- and four-spin interactions. A similar model has been studied
recently in the context of a quantum adiabatic algorithm\cite{peng08}
and a branch and bond optimization method.\cite{garc10}

Notice for this problem that the bond strengths are powers of two and
therefore cover an enormous range.  On the other hand, we also note
that the strongest bond approximately equals the sum of the other
bonds, $\sum_{i=0}^{n-1}2^i=2^n-1$, a limit to which we will return in
the section on greedy algorithms. Flipping all other spins in an
integer can therefore approximately cancel the effect of flipping the
highest spin.  As a simple example consider the drastic effect of
"carrying" in the following addition $01111111+1=10000000$, 
%which clearly demonstrates the lack of a simple measure of nearness in spin
%space for the above model.
which clearly demonstrates the lack of a simple connection between Hamming distance in spin space and the distance in
energy.  %SMG0202  Recall Nick Read didn't like this sentence.  Does this help?

Introducing a fictitious temperature $T=1/\beta$ the thermal
expectation value of the energy is given by $\langle H \rangle =\sum_i
E_ie^{-\beta E_i}/\sum_i e^{-\beta E_i},$ where the sum is over all
states, and $E_i$ is the energy of state $i$. It is worth noting that
the factorization problem differs from the spin glass problem in that
we know the lowest energy eigenvalue, namely zero, and we need to find
only the lowest energy eigenstate. During a stochastic simulation one
can therefore immediately interrupt the execution of the program when
the true ground state is found, which makes the factorization problem
an ideal testing case for ground state algorithms. We have generated
several instances of the factorization problem for increasing system
size, and in Table~\ref{tab:primes} we list the series of composite
integers, along with the prime factors, used in this work.

\begin{table}[htp]
  \begin{tabular}{|r|c|c|c|} \hline
    $n$ & $p_1$ & $p_2$ & $q$ \\ \hline\hline
    10 & 601 & 911 & 547511 \\
    12 & 2081 & 3329 & 6927649 \\
    14 & 10007 & 15091 & 151015637 \\
    16 & 40093 & 60013 & 2406101209 \\
    18 & 150011 & 140007 & 36003690077 \\
    20 & 700057 & 900001 & 630052000057 \\
    22 & 2500339 & 3500227 & 8751754076953 \\
    24 & 11600489 & 14000083 & 162407808840587 \\
    26 & 41615281 & 61616479 & 2564187087815599 \\
    28 & 150243361 & 220293523 & 33090127134000803 \\
    30 & 800000087 & 900000083 & 720000144700007221 \\
    \hline
\end{tabular}
\caption{Pregenerated composite integers along with factors.}
\label{tab:primes}
\end{table}

\section{Greedy algorithms}

The Edward-Anderson model, $H=\sum_{\langle ij\rangle} J_{ij}S_iS_j$,
is the archetypical Ising spin glass model.  The coupling parameters
$J_{ij}$ are quenched random variables, for example Gaussian
distributed with mean zero.  The combination of disorder and
frustration makes spin glass models very challenging to solve, and
there is no known general solution. However, in the limit of very
broadly distributed coupling constants $J_{ij}$ finding the ground
state of this model maps to a minimum spanning tree problem, solvable
by a greedy algorithm.\cite{newm94, jack10} The criterion is that the
magnitude of each coupling is greater than the absolute sum of all
smaller couplings, which, as a consequence, means that the width of
the distribution increases with system size. This may, at first, seem
to hold also for the factorization model, Eq.~(\ref{factor}), since
$\sum_{i=0}^{n-1}2^i=2^n-1 < 2^n$. However, there are two
caveats. First, all spin pairs for which $i+j=k$ share the same
coupling constant $J_k$, so there are many coupling constants with the
same magnitude. Second, when the model is formulated so that the prime
factors are given by the ground state, like $H_2$, the resulting model
contains multi-spin interactions.

Next we consider the limit of an even broader bond distribution and
show that it leads to important simplifications also for the
factorization model.  As mentioned above there are, in general, many
couplings of a given magnitude $J_k$. They all contribute to the
$k$:th digit of the composite integer $q$. In addition a carry bit
must be added. If we can avoid the carry bit the problem simplifies in
that the bonds of magnitude $J_k$ directly determine the $k$:th digit
of $q$.  This means that we can, in principle, satisfy the bonds (and
corresponding digits in q) in decreasing order without running the
risk of weaker bonds upsetting already satisfied stronger bonds.

As an illustration, we consider prime factors of the form
$p=\sum_{i=0}^{n-1} k^iS_i$, where $S\in\{0,1\}$, but $k>2$. If $k>n$
no carry operation is needed, and here we consider the case of
$k=10$. Examples of such prime factors include (in base 10
representation): 11, 101, 1011, 101111, 1011001, 1100101. The simplest
case is $11\times101=1111$.  Using long multiplication we work out the
factors simultaneously from the high (since there are no carry bits)
and low end:
$$
{\setlength\arrayrulewidth{.8pt}
\setlength\tabcolsep{4pt}
\begin{tabular}{*{5}{>{$}c<{$}}}
  && c & b & a\\
  \times &&f& e & d \\\hline
  &  & cd & bd & ad\\
    & ce & be & ae & \\
  cf & bf & af &  &\\\hline
   0 & 1 & 1 & 1 & 1
\end{tabular}}
$$
At the high end $cf=0$ and $ce+bf=1$ imply that c=1, f=0 (or vice
versa) and e=1. At the low end $ad=1$ implies that $a=d=1$. Both
remaining conditions yield b=1 rendering the factors 11 and
101. Solving the problem decade by decade is therefore much simpler in
the limit of a bond distribution broad enough to prevent the carry
operation. Thus, while the distribution of bond strengths in the full
factorization problem is exponentially broad, it is not broad enough
to allow us to apply a greedy algorithm in which we satisfy the bonds
one at a time.

\section{ Stochastic Algorithms}

Stochastic Monte Carlo methods have long been used to calculate
thermal expectation values, as well as ground state properties, of
statistical models. Depending on the model, and method, large systems
can often be studied to high precision. Using cluster algorithms the
critical temperature and other properties for three-dimensional
Ising-like models have been determined up to 6 decimals by performing
simulations of systems containing $128^3$ spins\cite{deng03}, and the
ground state energy of the two-dimensional Heisenberg model has been
determined to 5 decimals using quantum Monte Carlo
methods\cite{sand97, beard96}. There are also many examples in which
the use of efficient algorithms can change the functional dependence
of the scaling with respect to computational effort.  The use of
cluster updates at the critical point for the Ising model in effect
eliminates the problem of critical slowing down, and the computational
effort decreases from $L^{d+2}$ for single-spin updates to $L^d$  for
cluster updates, where $d$ is the dimensionality of the model and $L$
is the linear system size.\cite{new99} In some cases use of the loop
algorithms can make the exponentially difficult sign problem tractable
in polynomial time.\cite{chan99, hene00}

This suggests that Monte Carlo methods developed for statistical
physics problems could be useful in analyzing the factorization
problem, since finding factors containing a few hundred bits is
considered a hard problem.\cite{pome96} However, the Ising model that
arises from the factorization problem has a very complex energy
landscape, with multiple local minima separated by very high barriers.
Factorization is but one of many such difficult optimization problems
whose solution requires finding a global minimum in a very complex
landscape.  Other examples include various aspects of circuit design
in electronics,\cite{giel90} protein folding in life science\cite{pande2000},
spin-glass behavior in materials science\cite{youn97} and the
traveling salesman problem\cite{lawl85, hopf86} in computer science and
mathematics. In an attempt to alleviate the problems associated with
the complicated energy landscape the methods of thermal
annealing\cite{kirk83} and parallel tempering\cite{mari92} have been
developed. A fictitious or real temperature is introduced and as this
parameter is lowered, the system settles in a local minimum. If the
cooling is sufficiently slow the ground state is found.\cite{gema84}
However, for many complex systems it is practically impossible to
reduce the temperature slowly enough.

Classical simulated annealing relies on thermal fluctuations to find
the ground state. If the energy barriers separating different minima
are high and narrow, quantum tunneling can be a more efficient way to
equilibrate. Quantum mechanical systems are able to tunnel through
barriers, instead of traversing the barriers.  In quantum annealing
the minimum is found using the quantum mechanical tunneling effect
instead of thermal fluctuations. The efficiency of quantum annealing
has been studied both experimentally\cite{broo99} and
computationally\cite{farh01,sant02,hogg03,sant06, youn08, youn10}, in
which case quantum annealing can be realized by introducing
off-diagonal terms into the classical model, which cause tunneling
between the diagonal, classical states.

Next we consider a number of different approaches to a Monte Carlo
simulation of the factorization problem in order of increasing
complexity.

\subsection{Random search}

The simplest stochastic method for solving the factorization problem
is to generate random integers $p$ and interrupt the search when
$q\:\mbox{mod}\: p=0$.  The probability of finding a factor is $
\mu=2/2^n $ for a single attempt (since there are two factors). Since
the probability is constant for each attempt the number of trials
before success, $P(N)$, is Poisson distributed, $P(N)= \mu\exp(-\mu
N)$, with an average of $ \bar{N} = \mu^{-1}=2^{n-1} $.

\begin{figure}[h]
\resizebox{\hsize}{!}{\includegraphics{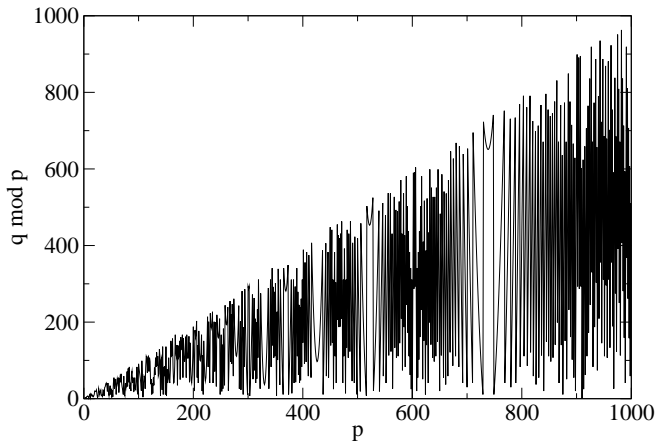}}
\caption{\label{fig:pd} The remainder after dividing q=547511 by
p. The remainder is equal to zero for p= 601 and 911}
\end{figure}

We use the random search as a point of reference and next we consider
Monte Carlo techniques using importance sampling. If the weight
function for a problem is fairly smooth and dominated by a few
pronounced maxima, then it may be possible to generate states distributed
according to the weight function, with a substantial gain in
performance. However, the weight function for the factorization
problem is very complex and the potential gain less certain. To
demonstrate the complicated structure of the factorization problem we
plot the function $q\:\mbox{mod}\:p$ for the case of
$q=547511=601\times 911$ in Fig.~\ref{fig:pd}
We note an average linear increase of $q\:\mbox{mod}\:p$, but
otherwise there is no apparent ordered structure. In the next sections
we investigate whether importance sampling can nevertheless still be
used to improve convergence to the ground state.

\subsection{Simple spin flips and  local temperature}
\begin{figure}[h]
\resizebox{\hsize}{!}{\includegraphics{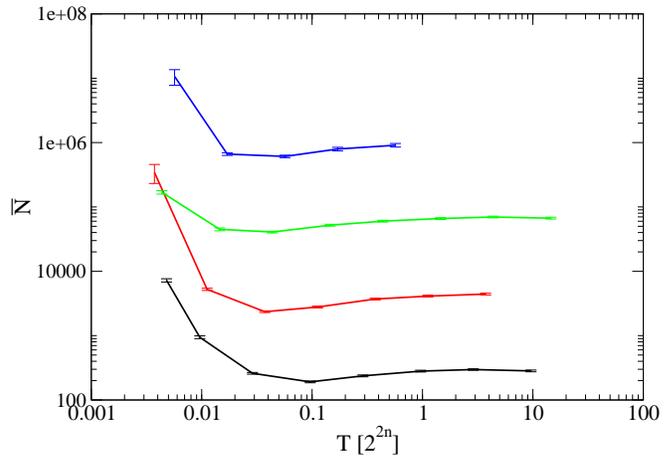}}
\caption{ The average number of attempts, $\bar{N}$ before finding a
  prime factor as a function of temperature. The temperature is shown
  in units of 2$^{2n}$. The curves are for $n$=22,18,14 and 10-spin
  factors from top to bottom.}
\label{metrop}
\end{figure}

Using the Boltzmann weight for state $i$, $W_i=\exp(-\beta E_i)$, we
implement a single-spin flip Metropolis algorithm for the model
defined by Eq.~(\ref{h1}). As a a single spin flip is attempted the
new weight function $W'$ is calculated, and the spin flip is accepted
with probability $p=\max (W'/W,1)= \max(\exp(-\beta \Delta E),1)$,
where $\Delta E=E'-E.$ To investigate the performance of the
Metropolis algorithm we start with two randomly chosen trial factors $p_1$
and $p_2$. During the simulation we choose spins at random and attempt
to flip single spins with the above probability. When one of the factors
is found the execution is halted, and the number of attempts, $N$ is
recorded. Note that all the states visited during the simulation are
counted, independently of whether the state is accepted or not. This is
repeated until a reliable average, $\bar{N}$ can be calculated
(typically one thousand runs). We restrict the search to odd
factors, by locking the lowest spin in the 1-state. This is to prevent
one of the factors becoming zero, in which case the energy is
independent of the value of the other factor. In Fig.~\ref{metrop} we
display the results obtained for some of the integers listed in
Tab.~\ref{tab:primes}.

\begin{figure}[h]
\resizebox{\hsize}{!}{\includegraphics{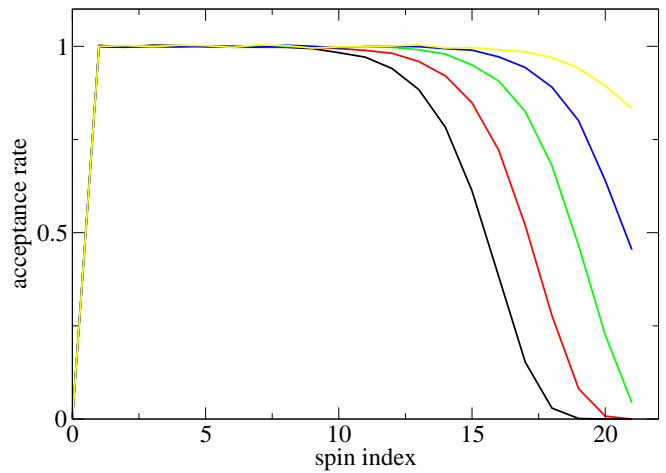}}
\caption{The acceptance rates for attempted spin flips of the 22-spin
  % factor. The temperature is 0.0056, 0.017, 0.057 and 0.56 in units
  % of 2$^{44}$ from left to right.  The temperature is $1\times
  % 10^{11}$, $3\times 10^{11}$, $1\times 10^{12}$, $3\times 10^{12}$,
  % $1\times 10^{13}$ from left to right.
}
\label{metrop_prob}
\end{figure}

For each system size there is a minimum at an intermediate
temperature, and as the temperature is lowered the number of attempts
increases dramatically, while it levels out at as the temperature is
increased. This behavior is understood considering
Fig.~\ref{metrop_prob}, where the acceptance probability for the
different spins are displayed for the five temperatures recorded in
Fig.~\ref{metrop} for n=22. At the highest temperature the acceptance
rate approaches unity for all the spins. This means that the spins are
freely fluctuating, and the results approach the random search
described above (consecutive configurations are still correlated, so
the random search is faster than the high temperature limit
of the Metropolis algorithm). As the temperature is lowered, the high
spins feel the effect first. For the nearest neighbor Ising model
the change in energy, $\Delta E=E'-E$ is of the order $J$, the uniform
coupling constant, for all spins considered. However, for the model
described by Eq.~(\ref{h1}), the change in energy is of order
$2^j\times p_2$ when an attempt is made to flip the $j$:th spin in
$p_1$. As the temperature is lowered, the probability of flipping the
high spins decreases quickly. The minimum in Fig.~\ref{metrop} occurs
when the probability of flipping the highest spin is about 0.05, just
before it freezes. Lowering the temperature further causes the number
of attempts to increase dramatically.

This indicates that one could adjust the model, given by
Eq.~(\ref{h1}) so that the probability distribution for accepting a
spin flip is more even. Any alterations are allowed as long as the
ground state is unchanged, and the aim is to decrease the level
spacing at higher energies. Therefore we consider the following forms
of the Hamiltonian,
\begin{equation}
H_{\frac{1}{2}}=\left|q-\sum_{i,j=1}^n2^{i+j}S^1_iS^2_j\right|^{\frac{1}{2}},
\label{h2}
\end{equation}
and
\begin{equation}
H_{\ln}=\ln\left(\left|q-\sum_{i,j=1}^n2^{i+j}S^1_iS^2_j\right|+1\right).
\label{h3}
\end{equation}
Both the square root and the logarithm are monotonically increasing
functions that do not change the order of the states. The logarithm
function, in particular, approximately cancels the exponential factor
$2^j\times p_2$ and allows all the spin to be updated in a more even
fashion.

\begin{figure}[h]
\resizebox{\hsize}{!}{\includegraphics{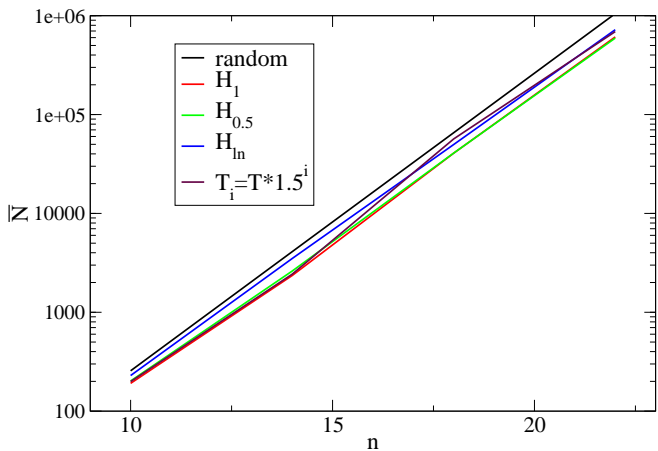}}
\caption{ The number of states visited before the $n$-spin prime
  factors are found for a random search, $H_1$ (absolute value),
  $H_{\frac{1}{2}}$ (square root), $H_{\ln}$ (logarithm) and local
  temperature $T_i=T\times 1.5^i$.}
\label{scaling}
\end{figure}

Yet another way to increase the fluctuations of the higher spins is to
introduce a site-dependent temperature, $T_i$.  A higher
temperature at the higher spins ensures more even fluctuations. We
therefore define a local temperature $T_i=T\times k^i$, with a
parameter $k$ whose value can vary from 1 (no change) to 2 (cancels
the exponential factor $2^j\times p_2$). In Fig.~\ref{scaling} we
compare the efficiency of the different approaches. We display the
number of states visited before a prime factor is found as a function
of system size. The temperature is adjusted for each data point to
optimize the search.  All methods are compared to the random search,
for which $\bar{N}=2^{n-2}$, since we restrict the search to odd
integers, and there are two distinct prime factors. We notice that all
methods represent slight improvements over the random search, but no
model works better than $H_1$ (absolute value),
which scales like $0.58\times 2^{n-2}$ and requires about half the
number of visited states compared to a random search.

\begin{figure}[h]
  \resizebox{\hsize}{!}{\includegraphics{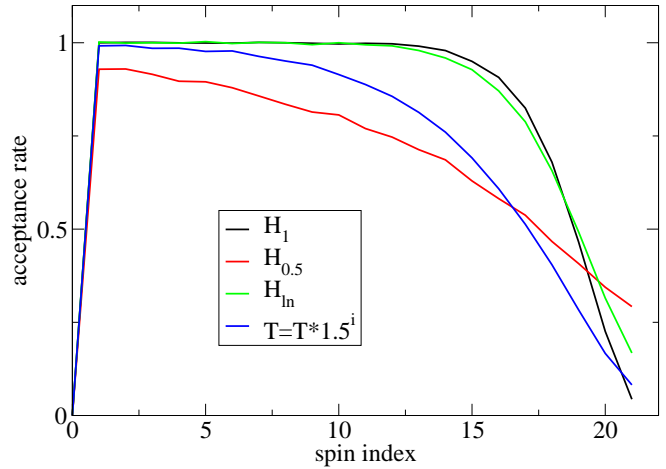}}
  \caption{ The acceptance rate for spin flips as a function of spin
    index for a 22-spin factor. Rates displayed for $H_1$ (absolute
    value), $H_{\frac{1}{2}}$ (square root), $H_{\ln}$ (logarithm) and
    local temperature $T_i=T\times 1.5^i$. }
\label{acc_prob}
\end{figure}

In Fig.~\ref{acc_prob} we compare the acceptance rates for the
different methods. The acceptance rate for $H_1$ (absolute value) and
$H_{\frac{1}{2}}$ (square root) quickly saturate to unity, while
$H_{\ln}$ (logarithm) and a local temperature scaling like
$T_i=T\times 1.5^i$ result in a much more even acceptance rate. The
results clearly show that obtaining a more even acceptance rate does
not necessarily speed up convergence to the ground state.

\subsection{Parallel tempering}

In the previous section the behavior of the model was investigated
while the temperature was held constant. More efficient algorithms for
converging to the ground state rely on a temperature that evolves
during the simulation. The main methods are annealing\cite{kirk83,
  gema84}, where the temperature is decreased as the simulation
progresses, and tempering methods\cite{mari92}, where the temperature
fluctuates between a maximum and a minimum during the simulation. Next
we implement a parallel tempering method for the factorization
problem.  In the factorization problem the energy scale and the
position of the spins are closely tied together. Once the energy has
been lowered sufficiently, the higher spins are entirely frozen, and
if they are not in the correct positions the ground state cannot be
found. Compared to annealing methods the tempering method offers the
advantage that the system can return to higher temperatures and
explore several local minima.

In the method of parallel tempering several copies, or replicas, of
the system are simulated concurrently. Each replica is initially
assigned a temperature, $T_i$, and after performing a number of Monte
Carlo updates at the assigned temperatures attempts are made to swap
nearby temperatures with a probability $P(T_i,T_j)=\exp
(E_i-E_j)/(T_i-T_j)$, which preserves detailed balance. The attempted
temperature swaps are repeated at regular intervals, and in this
manner the temperature of a given replica varies during the
simulation. The method has been very successful in equilibrating
disordered spin-glass systems at low temperatures as well as studying
phase transitions. Given the high energy barriers between low lying
states of the factorization model one could expect parallel tempering
to allow the replicas to transverse the barriers and not get stuck in
a given local minimum so easily.

\begin{figure}[h]
\resizebox{\hsize}{!}{\includegraphics{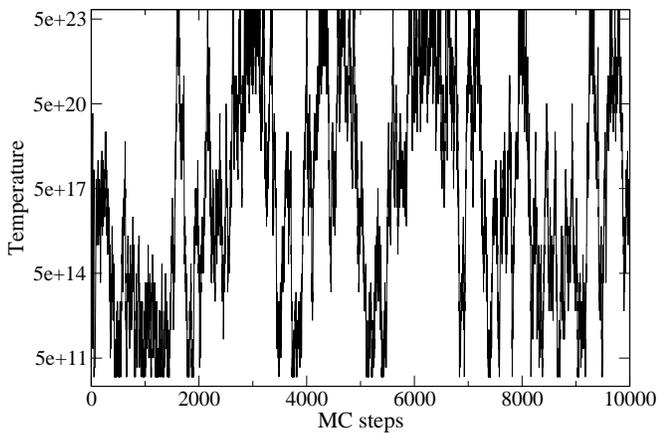}}
\caption{ The temperature fluctuations for replica during parallel
  tempering}
\label{pt}
\end{figure}

We have implemented a parallel tempering algorithm for the
factorization problem, based on the Metropolis algorithm of
$H_1$. One attempt is made, on average, to flip every spin in
all the replicas, and thereafter an attempt is made to switch all
neighboring temperatures. The maximum temperature is set to where the
Metropolis acceptance rate for flipping the highest spin is about 0.9,
and the lowest temperature when the acceptance rate for the lowest
spin is about 0.1. In Fig.~\ref{pt} we show the temperature
fluctuations for a single replica of a $n$=40-spin factorization
problem with $p_1= 100 000 000 0003$, $p_2=500 000 000 023$,
$q=p_1*p_2=50000000003800000000069$. The temperature varies from a
maximum of $T_{max}=1\times10^{24}$ to a minimum of
$T_{min}=1\times10^{11}$. In between there are 40 temperatures that
ensure that the swap rates remain above 0.5. As can be seen from the
figure the temperature of the replica wanders repeatedly back and
forth between the highest and lowest temperatures.

However, counting the total number of attempted spin updates in all
the replicas until a prime factor is encountered we find that the
method is not more efficient than the Metropolis algorithm considered
in the last section. This indicates that although the tempering method
improves equilibration at low and finite temperatures, it is not, in
this case, superior in picking out the ground state itself. There are
many low-lying energy states, and finding precisely the right one is a
difficult problem.

\subsection{Classical SSE cluster update}

So far we have discussed single spin flips, but one goal of this
investigation is to determine whether cluster updates, which have
proved highly useful for many difficult problems, can be of use for
the factorization problem. Since flipping the $j$:th spin in factor
$p_1$ changes the energy by $2^j\times p_2$ one could, in principle,
offset the large energy change by simultaneously flipping several
spins. Cluster updates for long-range classical models have been
developed within the framework of the Swendsen-Wang
update\cite{blot02} and the stochastic series
expansion\cite{sand03}. However, the factorization models lacks the
up-down symmetry of the standard Ising model, which cluster updates
usually rely on. If the spin variables assume the values $\pm 1$ the
product $S_iS_j$ is unchanged if both spins are flipped.  This is not
the case for the factorization model of Eq.~(\ref{h1}) since $S$ takes
the values 0 and 1.  By a transformation $S'=2S-1$ we can introduce a
new variable $S'$ that takes the values $\pm 1$, but this introduces
single spin operators $S'_i$ in Eq.~(\ref{h1}), which also destroy the
up-down symmetry.  Therefore we have not been able to introduce a
large-scale cluster update, but instead we implement a
``small-cluster'' update within the SSE method, which we describe
next.

The SSE method is based on a Taylor expansion of the partition
function Z,
$$ Z=\sum_{\alpha}\sum_{n=0}^{\infty} \frac{(-\beta)^n}{n!} \langle\alpha|H^n|\alpha\rangle,$$
where $|\alpha\rangle$ is a complete set of basis states. The SSE
method has been applied to Ising model with arbitrary interactions,
and we refer to Ref.~\onlinecite{sand03} for a detailed
description. Here we only describe modifications that arise when
applying the method to the factorization model.

Expressing the multiplication of two integers in binary form we obtain
the model
\begin{equation}
H=\sum_{i,j=1}^nJ_{i,j}S^1_iS^2_j,
\label{h5}
\end{equation}
with $J_{i,j}=2^{i+j}$.
Defining the bond operator $H_{i,j}=J_{i,j} (1-S^1_iS^2_j)$ this can
be written as
\begin{equation}
H=\sum_{i,j=1}^n-H_{i,j} + \sum_{i,j=1}^nJ_{i,j}.
\end{equation}
Including additional unit operators $I$, the Taylor expansion can be
written as
\begin{equation}
  Z=\frac{1}{L!}\sum_{\alpha}\sum_{n=0}^{\infty}\sum_{S_L} \frac{\beta^n(L-n)!}{L!}\langle  \alpha | S_L|\alpha\rangle,
\label{Z}
\end{equation}
where we have introduced a cut-off at $n=L$, and $S_L$ is the operator
string $S_L=\prod_{p=1}^L H_p$ with $H_p\in \{H_{i,j},I\}$. The matrix
element in Eq.~(\ref{Z}) can be written as a product of elements of
the form
$$\langle S^1_i S^2_j| H_{i,j} | S^1_iS^2_j\rangle,$$
and for the model we consider only the matrix elements $\langle 0 0
| H_{i,j}| 00\rangle = \langle 0 1 | H_{i,j}| 01\rangle=\langle 1 0 |
H_{i,j}| 10\rangle=J_{ij}$ contribute since $\langle 1 1 |
H_{i,j}| 11 \rangle =0$.

In order to sample the configuration space of all operator sequences
$S_L$ and all states $|\alpha\rangle$, two types of updates are
necessary. The first update changes the number of non-identity
operators in the sequence by attempting to exchange identity and bond
operators. The probability of exchanging a unit operator for a bond
operator is
$$P_{bond}= \frac{\beta\sum_{i,j}J_{ij}}{L-n+\beta\sum_{i,j}J_{ij}} $$
and the reverse probability of exchanging a bond operator for a unit
operator is
$$P_{unit}= \frac{L-n+1}{L-n+1+\beta\sum_{i,j}J_{ij}} $$
If a bond operator is to be inserted, the particular bond is chosen
according to the relative weight $J_{ij}$ of the bond, as described in
Ref.~\onlinecite{sand03}.

The update described above only changes the operator sequence $S_L$,
and the state $|\alpha\rangle $ is not affected. The classical cluster
update described in Ref.~\onlinecite{sand03} flips all the spins that
are interconnected by bond operators; since the Ising operators depend
on the relative orientation of the spins this does not change the
weight and is always allowed. For the factorization model flipping two
spins in the state $|0\rangle$ connected by a bond leads to the
forbidden $\langle 1 1| H_{ij}|11\rangle =0$ vertex.  Hence a
different update is needed, and so we implement a "small-cluster" move.

In the small-cluster move a spin is chosen at random and an attempt is
made to flip it. First we consider the case of a 1-spin. The spin in
question may be connected to other spins (in the other integer)
through bond operators. Let us denote these spins nearest-neighbor
(nn) spins.  The nn-spins, in turn, may be connected to spins in the
same integer as the spin we are attempting to flip. We call these
spins next nearest neighbor (nnn) spins. An attempt to flip a 1-spin
is always accepted, but in order to satisfy detailed balance with the
reverse update, to be described below, we also need to consider all nn
spins. If there is a nn spin connected to nnn spins, all of which take
the value 0, then these nn spins are assigned values 0 and 1 with
equal probability as the original spin is flipped.

\begin{figure}[h]
\resizebox{\hsize}{!}{\includegraphics{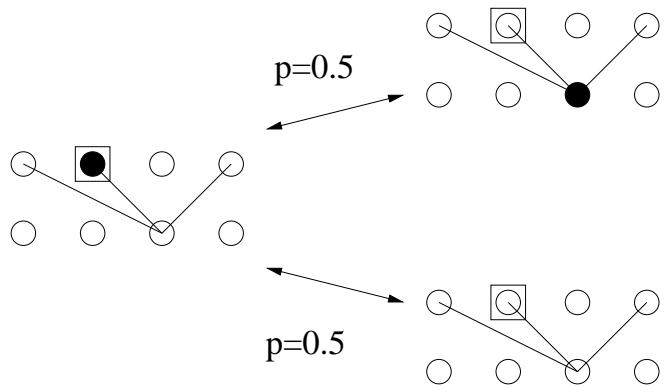}}
\caption{ Illustration of a small-cluster update satisfying detailed
  balance. The upper row of spins denote one integer and the lower row
  the other integer. Spins in the 1 state are denoted by a filled
  circle, and spin spins in the 0-state be an empty circle. The spin
  marked by a square box is flipped between the 1-state (left side)
  and 0-state (right side). }
\label{sse}
\end{figure}

The reverse move is flipping a 0-spin to a 1-spin. This move is
accepted with probability $2^{-nn_0}$, where $nn_0$ is the number of
nn spins connected to nnn spins, all of which necessarily assume the
value 0 . If the move is accepted all the nn spins are set to 0. The
factor $2^{-nn_0}$ corresponds to the number of ways the nn spins can
be assigned states 0 and 1 in the above reverse move and ensures that
detailed balance is satisfied. An illustration of this move (with
$nn_0=1$) is shown in Fig.~\ref{sse}. The advantage of this update,
compared to a single-spin flip, is that is allows 0-spins to be
updated even though they may be connected to 1-spins.

The two updates together ensure ergodicity, satisfy detailed balance
and demonstrate that it is possible to use an update procedure that
flips more than one spin at a time. However, it only works for the
model Eq.~(\ref{h5}), which does not have the prescribed integer as a
ground state. One way to still use the small-cluster update is to
allow only updates that do not lead to a state with an energy lower
than the prescribed composite integer q. This we have implemented, but
unfortunately the scaling, once again, is not better than simple
Metropolis updates.

\subsection{Transverse Field}

\begin{figure}[h]
\resizebox{\hsize}{!}{\includegraphics{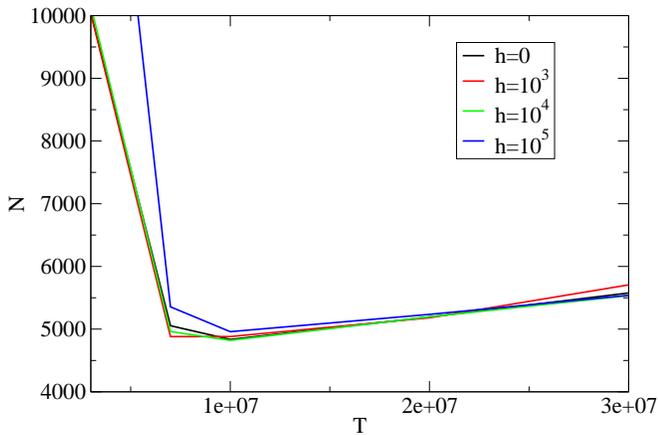}}
\caption{ The average number of attempts, $\bar{N}$ before finding a
  prime factor as a function of temperature and transverse field for a
  14-spin factor.}
\label{tf}
\end{figure}

The above algorithms are based on thermal fluctuations. However, it is
also possible to modify the model and introduce quantum mechanical
terms. Since quantum mechanical systems are able to penetrate through
energy barriers, instead of going over energy barriers this method can,
in certain cases be superior. We have therefore added a a transverse
magnetic field to the model,
\begin{equation}
H_3=\left|p-\sum_{i,j}2^{i+j}S_{1,i}S_{2,j}\right|+
h\sum_i (S_i^+ + S_i^-),
\label{eq:trans}
\end{equation}
in order to compare the efficiency of the quantum and thermal
fluctuations. First we consider a transverse field that is constant in
time, and find the temperature and field strength combination that
minimizes the number of states visited before finding the ground
state. We implement the transverse field using a continuous time
algorithm\cite{beard96, rieg99}. Individual spins now fluctuate in
imaginary time, $S(\tau)$, and imaginary-time segments of individual
spins can be flipped. However, due to the difficulties explained in
the above section we do not use a cluster update, but an
imaginary-time spin segment $\{\tau_0,\tau_1\}$ is flipped with
probability $\exp(\int_{\tau_0}^{\tau_1}d\tau\tau\Delta E(\tau)$.

In Fig.~\ref{tf} we show the effect of increasing the transverse field
for the $N=14$ spin system.  Interestingly it appears that, in this
case, the quantum fluctuations are not more efficient in finding the
ground state than thermal fluctuations. The smallest number of steps
are found for the case of zero transverse field. We have also used an
exponentially increasing field, $h_i=1.5^i,$ which leads to more
fluctuations in imaginary time for the higher spins, but we find that
this does not improve the scaling.

\subsection{Quantum annealing}
\begin{figure}[h]
\resizebox{\hsize}{!}{\includegraphics{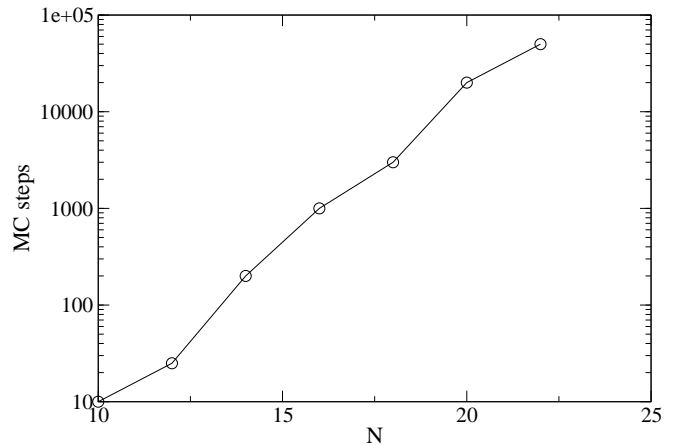}}
\caption{ The complexity, or number of MC steps, necessary to find the
  correct prime factors using an imaginary time quantum annealing
  method.}
\label{qa}
\end{figure}
While in the last subsection we considered the effect of a
time-independent transverse field it is more common to gradually
reduce the quantum mechanical terms during the simulation. A commonly
used annealing method\cite{farh01} defines a Hamiltonian
\begin{equation} H=sH_c +(s-1)H_q,
\label{eq:qa}
\end{equation}
where $H_c$ is the classical model with the desired ground
state. Quantum fluctuations are introduced through $H_q$, which, for
Ising systems, usually is a transverse field. The control parameter
$s$ is slowly reduced from the initial value $s=0$ to the final value
$s=1$ as the system evolves according to the time-dependent
Schr\"odinger equation
$$ i\hbar \frac{d}{dt} |\psi(t)\rangle=H(t)|\psi (t)\rangle.$$
The process is performed at a very low temperature, and, if the
process is sufficiently slow as to be adiabatic, the system evolves from
the ground state of $H_q$ to the ground state of $H_c$, which is the
solution to the problem. The time required to find the correct ground
state with a significant probability is called the complexity of the
problem, and numerical simulations of small systems indicate that the
complexity may (in some cases) increase polynomially in system
size.\cite{farh01,hogg03}

\begin{figure}[h]
\resizebox{\hsize}{!}{\includegraphics{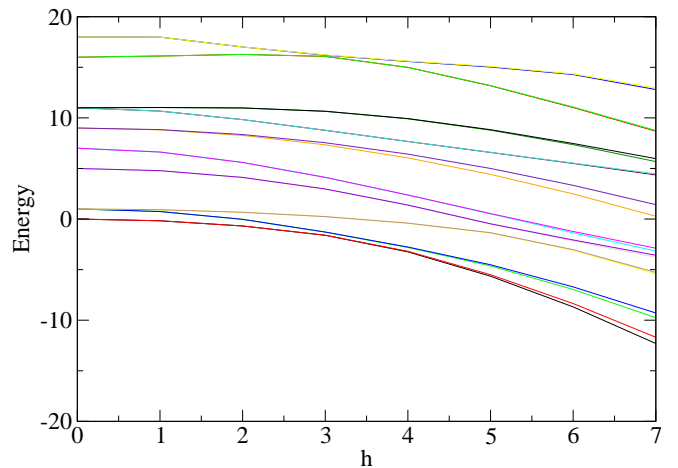}}
\caption{ The 20 lowest energy levels as function of transverse field
  for an instance of the factorization problem (q=551) with two five
  bit integers.}
\label{level}
\end{figure}

Real-time quantum annealing governed by the Schr\"odinger equation is
limited by the size of the minimal gap between the ground state and
first excited state. In Fig.~\ref{level}, the 20 lowest energy levels
for the model defined by Eq.~(\ref{eq:trans}) are displayed. The
system consist of two five bit integers with $q=551$, and we do
observe a minimum in the gap around $h=3$. It appears that, at least
in some random satisfiability problems, this gap is exponentially
small for certain instances of the problems due to a first-order phase
transition.\cite{youn08,youn10} This effect could limit the
applicability of quantum annealing methods to smaller system sizes.  A
recent study shows that, for small system sizes, real-time quantum
annealing of the factorization model does indeed scale polynomially in
system size.\cite{peng08} In order to find out whether the scaling
persists to larger system sizes it would be necessary to study the
scaling of the ground state energy gap with system
size.\cite{youn08,youn10}

Here we only use imaginary-time dynamics to study larger system sizes.
This does not constitute the real time dynamics of the Schr\"odinger
equation, but limited success has nevertheless been
reported.\cite{sant06} In order to test this method on the
factorization problem we study the model
\begin{equation}
  H= s\left| q- \sum_{i,j=1}^n2^{i+j}S^1_iS^2_j\right| + (s-1)h \sum_i (S_i^+ + S_i^-).
\end{equation}
The temperature $T$ is set so that virtually no spin flips are
accepted when s=1, and the strength of the transverse field is set to
$h=10\:T$. We have determined the number of sweeps of the whole
lattice (MC steps) necessary to find the correct ground state with a
probability of about 30$\%$, and display the result in
Fig.~\ref{qa}. From the figure we see that the system size dependence is
still exponential, and it appears that the standard
quantum annealing method as applied in an imaginary-time path-integral
simulation does not improve the scaling.

\section{Discussion and Conclusion}

We have developed a statistical mechanics Monte Carlo approach to the
factorization problem.  The resulting model is highly complex with
exponentially large frustrated long-range multi-spin interactions.  We
found that importance sampling is only very weakly effective in
improving the convergence to the ground state.  The fastest method we
found remains exponential and beats random sampling by only a factor
two.

We believe that the challenge to standard statistical methods is
threefold.  First, as for other difficult optimization problems, there
is a complex energy landscape with many low-lying states separated by
high energy barriers. In this case the global minimum is required and
therefore tempering and annealing methods that have been so successful
in determining the low-temperature properties of spin glasses are of
only limited use.

Second, unlike in standard short-range spin glass models the energy
change resulting from a single spin update has an exponentially broad
distribution, implying that most single-spin acceptance rates are
either close to unity or close to zero. This gives the importance
sampling the character of a random search.  Rescaling the temperature
and transverse fields for individual spins did not ameliorate this
problem.  We also found that while the distribution of bond strengths
was exponentially broad with a width increasing with system size, it
was not broad enough to permit solution via a greedy algorithm.

Third, as a direct consequence of the broadly distributed energy
change following a single spin flip it follows that there is no
concept of nearness in spin space. Two states with nearly identical
spin configurations may have vastly different energies. Standard
importance sampling methods are based on small changes in energy, a
requirement not easily satisfied for the factorization problem.

The statistical physics model resulting from the integer factorization
problem is a good benchmark model for ground state algorithms since
the ground state energy is known by construction. We hope that this
work may encourage further investigations in the efficiency of
possible cluster algorithms and annealing methods for the integer
factorization problem. All methods considered in this work require
$\mathcal{O}(q^\frac{1}{2})$ operations, to factor a composite
integer $q$. Whether it is possible to find a stochastic method that
scales with an exponent less than $ \frac{1}{2}$ remains an open
question. Another interesting question which we have not addressed is
the existence of a finite-temperature spin-glass transition. The
factorization model is frustrated and, to some extent,
disordered. These are considered two necessary conditions for the
existence of a stable spin-glass phase. A freezing of the spins would
certainly protect the ground state and further explain the difficulty
of solving the factorization problem.

\begin{acknowledgments}
We thank Lev Bishop, Nicholas Read and Daniel Spielman for helpful conversations.
  P.H. acknowledges support by the Swedish Research Council and the
  G\"oran Gustafsson Foundation. S.M.G. acknowledges support by NSF
  DMR-1004406, DMR-0653377.
\end{acknowledgments}

\bibliography{factor}

\begin{thebibliography}{35}
\expandafter\ifx\csname natexlab\endcsname\relax\def\natexlab#1{#1}\fi
\expandafter\ifx\csname bibnamefont\endcsname\relax
  \def\bibnamefont#1{#1}\fi
\expandafter\ifx\csname bibfnamefont\endcsname\relax
  \def\bibfnamefont#1{#1}\fi
\expandafter\ifx\csname citenamefont\endcsname\relax
  \def\citenamefont#1{#1}\fi
\expandafter\ifx\csname url\endcsname\relax
  \def\url#1{\texttt{#1}}\fi
\expandafter\ifx\csname urlprefix\endcsname\relax\def\urlprefix{URL }\fi
\providecommand{\bibinfo}[2]{#2}
\providecommand{\eprint}[2][]{\url{#2}}

\bibitem[{\citenamefont{Shor}(1997)}]{shor97}
\bibinfo{author}{\bibfnamefont{P.~W.} \bibnamefont{Shor}},
  \bibinfo{journal}{SIAM J. Comput.} \textbf{\bibinfo{volume}{26}},
  \bibinfo{pages}{1484} (\bibinfo{year}{1997}).

\bibitem[{\citenamefont{Lenstra and JR.}(1993)}]{lens93}
\bibinfo{editor}{\bibfnamefont{A.~K.} \bibnamefont{Lenstra}} \bibnamefont{and}
  \bibinfo{editor}{\bibfnamefont{H.~W.~L.} \bibnamefont{JR.}}, eds.,
  \emph{\bibinfo{title}{The development of the number field sieve}}, vol.
  \bibinfo{volume}{1554} of \emph{\bibinfo{series}{Lecture Notes in Math.}}
  (\bibinfo{publisher}{Springer-Verlag, Berlin and Heidelberg},
  \bibinfo{year}{1993}).

\bibitem[{\citenamefont{Pomerance}(1996)}]{pome96}
\bibinfo{author}{\bibfnamefont{C.}~\bibnamefont{Pomerance}},
  \bibinfo{journal}{Notices of the AMS} \textbf{\bibinfo{volume}{43}},
  \bibinfo{pages}{1473} (\bibinfo{year}{1996}).

\bibitem[{\citenamefont{Peng et~al.}(2008)\citenamefont{Peng, Liao, Xu, Qin,
  Zhou, Suter, and Du}}]{peng08}
\bibinfo{author}{\bibfnamefont{X.}~\bibnamefont{Peng}},
  \bibinfo{author}{\bibfnamefont{Z.}~\bibnamefont{Liao}},
  \bibinfo{author}{\bibfnamefont{N.}~\bibnamefont{Xu}},
  \bibinfo{author}{\bibfnamefont{G.}~\bibnamefont{Qin}},
  \bibinfo{author}{\bibfnamefont{X.}~\bibnamefont{Zhou}},
  \bibinfo{author}{\bibfnamefont{D.}~\bibnamefont{Suter}}, \bibnamefont{and}
  \bibinfo{author}{\bibfnamefont{J.}~\bibnamefont{Du}}, \bibinfo{journal}{Phys.
  Rev. Lett.} \textbf{\bibinfo{volume}{101}}, \bibinfo{pages}{220405}
  (\bibinfo{year}{2008}).

\bibitem[{\citenamefont{Garcia and Markov}(2010)}]{garc10}
\bibinfo{author}{\bibfnamefont{H.}~\bibnamefont{Garcia}} \bibnamefont{and}
  \bibinfo{author}{\bibfnamefont{I.}~\bibnamefont{Markov}}
  (\bibinfo{year}{2010}), \eprint{arXiv:0912.3912}.

\bibitem[{\citenamefont{Kruskal}(1956)}]{krusk56}
\bibinfo{author}{\bibfnamefont{J.~B.} \bibnamefont{Kruskal}},
  \bibinfo{journal}{Proc. Am. Math. Soc.} \textbf{\bibinfo{volume}{7}},
  \bibinfo{pages}{48} (\bibinfo{year}{1956}).

\bibitem[{\citenamefont{Read}(2005)}]{read05}
\bibinfo{author}{\bibfnamefont{N.}~\bibnamefont{Read}}, \bibinfo{journal}{Phys.
  Rev. E.} \textbf{\bibinfo{volume}{72}}, \bibinfo{pages}{036114}
  (\bibinfo{year}{2005}).

\bibitem[{\citenamefont{Newman and Stein}(1994)}]{newm94}
\bibinfo{author}{\bibfnamefont{C.~M.} \bibnamefont{Newman}} \bibnamefont{and}
  \bibinfo{author}{\bibfnamefont{D.~L.} \bibnamefont{Stein}},
  \bibinfo{journal}{Phys. Rev. Lett.} \textbf{\bibinfo{volume}{72}},
  \bibinfo{pages}{2286} (\bibinfo{year}{1994}).

\bibitem[{\citenamefont{Jackson and Read}(2010)}]{jack10}
\bibinfo{author}{\bibfnamefont{T.~S.} \bibnamefont{Jackson}} \bibnamefont{and}
  \bibinfo{author}{\bibfnamefont{N.}~\bibnamefont{Read}},
  \bibinfo{journal}{Phys. Rev. E.} \textbf{\bibinfo{volume}{81}},
  \bibinfo{pages}{021130} (\bibinfo{year}{2010}).

\bibitem[{\citenamefont{Bhatt and Lee}(1982)}]{bhat82}
\bibinfo{author}{\bibfnamefont{R.~N.} \bibnamefont{Bhatt}} \bibnamefont{and}
  \bibinfo{author}{\bibfnamefont{P.~A.} \bibnamefont{Lee}},
  \bibinfo{journal}{Phys. Rev. Lett.} \textbf{\bibinfo{volume}{48}},
  \bibinfo{pages}{344} (\bibinfo{year}{1982}).

\bibitem[{\citenamefont{Fisher}(1994)}]{fish94}
\bibinfo{author}{\bibfnamefont{D.~S.} \bibnamefont{Fisher}},
  \bibinfo{journal}{Phys. Rev. B.} \textbf{\bibinfo{volume}{50}},
  \bibinfo{pages}{3799} (\bibinfo{year}{1994}).

\bibitem[{\citenamefont{Deng and Bl{\"o}te}(2003)}]{deng03}
\bibinfo{author}{\bibfnamefont{Y.}~\bibnamefont{Deng}} \bibnamefont{and}
  \bibinfo{author}{\bibfnamefont{H.~W.~J.} \bibnamefont{Bl{\"o}te}},
  \bibinfo{journal}{Phys. Rev E.} \textbf{\bibinfo{volume}{68}},
  \bibinfo{pages}{036125} (\bibinfo{year}{2003}).

\bibitem[{\citenamefont{Sandvik}(1997)}]{sand97}
\bibinfo{author}{\bibfnamefont{A.}~\bibnamefont{Sandvik}},
  \bibinfo{journal}{Phys. Rev B.} \textbf{\bibinfo{volume}{56}},
  \bibinfo{pages}{11678} (\bibinfo{year}{1997}).

\bibitem[{\citenamefont{Beard and Wiese}(1996)}]{beard96}
\bibinfo{author}{\bibfnamefont{B.~B.} \bibnamefont{Beard}} \bibnamefont{and}
  \bibinfo{author}{\bibfnamefont{U.-J.} \bibnamefont{Wiese}},
  \bibinfo{journal}{Phys. Rev. Lett.} \textbf{\bibinfo{volume}{77}},
  \bibinfo{pages}{5130} (\bibinfo{year}{1996}).

\bibitem[{\citenamefont{Newman and Barkema}(1999)}]{new99}
\bibinfo{author}{\bibfnamefont{M.~E.~J.} \bibnamefont{Newman}}
  \bibnamefont{and} \bibinfo{author}{\bibfnamefont{G.~T.}
  \bibnamefont{Barkema}}, \emph{\bibinfo{title}{Monte Carlo Methods in
  Statistical Physics}} (\bibinfo{publisher}{Oxford University Press},
  \bibinfo{year}{1999}).

\bibitem[{\citenamefont{Chandrasekharan and Wiese}(1999)}]{chan99}
\bibinfo{author}{\bibfnamefont{S.}~\bibnamefont{Chandrasekharan}}
  \bibnamefont{and} \bibinfo{author}{\bibfnamefont{U.-J.} \bibnamefont{Wiese}},
  \bibinfo{journal}{Phys. Rev. Lett.} \textbf{\bibinfo{volume}{83}},
  \bibinfo{pages}{3116} (\bibinfo{year}{1999}).

\bibitem[{\citenamefont{Henelius and Sandvik}(2000)}]{hene00}
\bibinfo{author}{\bibfnamefont{P.}~\bibnamefont{Henelius}} \bibnamefont{and}
  \bibinfo{author}{\bibfnamefont{A.~W.} \bibnamefont{Sandvik}},
  \bibinfo{journal}{Phys. Rev. B} \textbf{\bibinfo{volume}{62}},
  \bibinfo{pages}{1102} (\bibinfo{year}{2000}).

\bibitem[{\citenamefont{Gielen et~al.}(1990)\citenamefont{Gielen, Walscharts,
  and Sansen}}]{giel90}
\bibinfo{author}{\bibfnamefont{G.~G.~E.} \bibnamefont{Gielen}},
  \bibinfo{author}{\bibfnamefont{H.~C.~C.} \bibnamefont{Walscharts}},
  \bibnamefont{and} \bibinfo{author}{\bibfnamefont{W.~M.~C.}
  \bibnamefont{Sansen}}, \bibinfo{journal}{IEEE J. Solid-State Circuits}
  \textbf{\bibinfo{volume}{25}}, \bibinfo{pages}{707} (\bibinfo{year}{1990}).

\bibitem[{\citenamefont{Pande et~al.}(2000)\citenamefont{Pande, Grosberg, and
  Tanaka}}]{pande2000}
\bibinfo{author}{\bibfnamefont{V.~S.} \bibnamefont{Pande}},
  \bibinfo{author}{\bibfnamefont{A.~Y.} \bibnamefont{Grosberg}},
  \bibnamefont{and} \bibinfo{author}{\bibfnamefont{T.}~\bibnamefont{Tanaka}},
  \bibinfo{journal}{Rev. Mod. Phys.} \textbf{\bibinfo{volume}{72}},
  \bibinfo{pages}{259} (\bibinfo{year}{2000}).

\bibitem[{\citenamefont{Young}(1997)}]{youn97}
\bibinfo{editor}{\bibfnamefont{A.~P.} \bibnamefont{Young}}, ed.,
  \emph{\bibinfo{title}{Spin Glasses and Random Fields}}
  (\bibinfo{publisher}{World Scientific}, \bibinfo{year}{1997}).

\bibitem[{\citenamefont{Lawler et~al.}(1985)\citenamefont{Lawler, Lenstra, Kan,
  and Shmoys}}]{lawl85}
\bibinfo{author}{\bibfnamefont{E.~L.} \bibnamefont{Lawler}},
  \bibinfo{author}{\bibfnamefont{J.~K.} \bibnamefont{Lenstra}},
  \bibinfo{author}{\bibfnamefont{A.~H. G.~R.} \bibnamefont{Kan}},
  \bibnamefont{and} \bibinfo{author}{\bibfnamefont{D.~B.}
  \bibnamefont{Shmoys}}, \emph{\bibinfo{title}{The Traveling Salesman Problem:
  A Guided Tour of Combinatorial Optimization}} (\bibinfo{publisher}{John Wiley
  \& Sons}, \bibinfo{year}{1985}).

\bibitem[{\citenamefont{Hopfield and Tank}(1986)}]{hopf86}
\bibinfo{author}{\bibfnamefont{J.~J.} \bibnamefont{Hopfield}} \bibnamefont{and}
  \bibinfo{author}{\bibfnamefont{D.~W.} \bibnamefont{Tank}},
  \bibinfo{journal}{Science} \textbf{\bibinfo{volume}{233}},
  \bibinfo{pages}{625} (\bibinfo{year}{1986}).

\bibitem[{\citenamefont{Kirkpatrick et~al.}(1983)\citenamefont{Kirkpatrick, Jr,
  and Vecchi}}]{kirk83}
\bibinfo{author}{\bibfnamefont{S.}~\bibnamefont{Kirkpatrick}},
  \bibinfo{author}{\bibfnamefont{C.~D.~G.} \bibnamefont{Jr}}, \bibnamefont{and}
  \bibinfo{author}{\bibfnamefont{M.~P.} \bibnamefont{Vecchi}},
  \bibinfo{journal}{Science} \textbf{\bibinfo{volume}{220}},
  \bibinfo{pages}{671} (\bibinfo{year}{1983}).

\bibitem[{\citenamefont{Marinari and Parisi}(1992)}]{mari92}
\bibinfo{author}{\bibfnamefont{E.}~\bibnamefont{Marinari}} \bibnamefont{and}
  \bibinfo{author}{\bibfnamefont{G.}~\bibnamefont{Parisi}},
  \bibinfo{journal}{Europhys. Lett.} \textbf{\bibinfo{volume}{19}},
  \bibinfo{pages}{451} (\bibinfo{year}{1992}).

\bibitem[{\citenamefont{Geman and Geman}(1984)}]{gema84}
\bibinfo{author}{\bibfnamefont{S.}~\bibnamefont{Geman}} \bibnamefont{and}
  \bibinfo{author}{\bibfnamefont{D.}~\bibnamefont{Geman}},
  \bibinfo{journal}{IEEE Trans. PAMI} \textbf{\bibinfo{volume}{6}},
  \bibinfo{pages}{721} (\bibinfo{year}{1984}).

\bibitem[{\citenamefont{Brooke et~al.}(1999)\citenamefont{Brooke, Bitko,
  Rosenbaum, and Aeppli}}]{broo99}
\bibinfo{author}{\bibfnamefont{J.}~\bibnamefont{Brooke}},
  \bibinfo{author}{\bibfnamefont{D.}~\bibnamefont{Bitko}},
  \bibinfo{author}{\bibfnamefont{T.~F.} \bibnamefont{Rosenbaum}},
  \bibnamefont{and} \bibinfo{author}{\bibfnamefont{G.}~\bibnamefont{Aeppli}},
  \bibinfo{journal}{Science} \textbf{\bibinfo{volume}{284}},
  \bibinfo{pages}{779} (\bibinfo{year}{1999}).

\bibitem[{\citenamefont{Farhi et~al.}(2001)\citenamefont{Farhi, Goldstone,
  Gutmann, Lapan, Lundgren, and Preda}}]{farh01}
\bibinfo{author}{\bibfnamefont{E.}~\bibnamefont{Farhi}},
  \bibinfo{author}{\bibfnamefont{J.}~\bibnamefont{Goldstone}},
  \bibinfo{author}{\bibfnamefont{S.}~\bibnamefont{Gutmann}},
  \bibinfo{author}{\bibfnamefont{J.}~\bibnamefont{Lapan}},
  \bibinfo{author}{\bibfnamefont{A.}~\bibnamefont{Lundgren}}, \bibnamefont{and}
  \bibinfo{author}{\bibfnamefont{D.}~\bibnamefont{Preda}},
  \bibinfo{journal}{Science} \textbf{\bibinfo{volume}{292}},
  \bibinfo{pages}{472} (\bibinfo{year}{2001}).

\bibitem[{\citenamefont{Santoro et~al.}(2002)\citenamefont{Santoro, Martonak,
  Tosatti, and Car}}]{sant02}
\bibinfo{author}{\bibfnamefont{G.~E.} \bibnamefont{Santoro}},
  \bibinfo{author}{\bibfnamefont{R.}~\bibnamefont{Martonak}},
  \bibinfo{author}{\bibfnamefont{E.}~\bibnamefont{Tosatti}}, \bibnamefont{and}
  \bibinfo{author}{\bibfnamefont{R.}~\bibnamefont{Car}},
  \bibinfo{journal}{Science} \textbf{\bibinfo{volume}{295}},
  \bibinfo{pages}{2427} (\bibinfo{year}{2002}).

\bibitem[{\citenamefont{Hogg}(2003)}]{hogg03}
\bibinfo{author}{\bibfnamefont{T.}~\bibnamefont{Hogg}}, \bibinfo{journal}{Phys.
  Rev. A.} \textbf{\bibinfo{volume}{67}}, \bibinfo{pages}{022314}
  (\bibinfo{year}{2003}).

\bibitem[{\citenamefont{Santoro and Tosatti}(2006)}]{sant06}
\bibinfo{author}{\bibfnamefont{G.}~\bibnamefont{Santoro}} \bibnamefont{and}
  \bibinfo{author}{\bibfnamefont{E.}~\bibnamefont{Tosatti}},
  \bibinfo{journal}{J. Phys. A.} \textbf{\bibinfo{volume}{39}},
  \bibinfo{pages}{R393} (\bibinfo{year}{2006}).

\bibitem[{\citenamefont{Young et~al.}(2008{\natexlab{a}})\citenamefont{Young,
  Knysh, and Smelyanskiy}}]{youn08}
\bibinfo{author}{\bibfnamefont{A.~P.} \bibnamefont{Young}},
  \bibinfo{author}{\bibfnamefont{S.}~\bibnamefont{Knysh}}, \bibnamefont{and}
  \bibinfo{author}{\bibfnamefont{V.~N.} \bibnamefont{Smelyanskiy}},
  \bibinfo{journal}{Phys. Rev. Lett.} \textbf{\bibinfo{volume}{101}},
  \bibinfo{pages}{170503} (\bibinfo{year}{2008}{\natexlab{a}}).

\bibitem[{\citenamefont{Young et~al.}(2008{\natexlab{b}})\citenamefont{Young,
  Knysh, and Smelyanskiy}}]{youn10}
\bibinfo{author}{\bibfnamefont{A.~P.} \bibnamefont{Young}},
  \bibinfo{author}{\bibfnamefont{S.}~\bibnamefont{Knysh}}, \bibnamefont{and}
  \bibinfo{author}{\bibfnamefont{V.~N.} \bibnamefont{Smelyanskiy}},
  \bibinfo{journal}{Phys. Rev. Lett.} \textbf{\bibinfo{volume}{104}},
  \bibinfo{pages}{020502} (\bibinfo{year}{2008}{\natexlab{b}}).

\bibitem[{\citenamefont{Bl{\"o}te et~al.}(2002)\citenamefont{Bl{\"o}te,
  Heringa, and Luijten}}]{blot02}
\bibinfo{author}{\bibfnamefont{H.}~\bibnamefont{Bl{\"o}te}},
  \bibinfo{author}{\bibfnamefont{J.}~\bibnamefont{Heringa}}, \bibnamefont{and}
  \bibinfo{author}{\bibfnamefont{E.}~\bibnamefont{Luijten}},
  \bibinfo{journal}{Comp. Phys. Comm.} \textbf{\bibinfo{volume}{147}},
  \bibinfo{pages}{58} (\bibinfo{year}{2002}).

\bibitem[{\citenamefont{Sandvik}(2003)}]{sand03}
\bibinfo{author}{\bibfnamefont{A.}~\bibnamefont{Sandvik}},
  \bibinfo{journal}{Phys. Rev. E} \textbf{\bibinfo{volume}{68}},
  \bibinfo{pages}{056701} (\bibinfo{year}{2003}).

\bibitem[{\citenamefont{Rieger and Kawashima}(1999)}]{rieg99}
\bibinfo{author}{\bibfnamefont{H.}~\bibnamefont{Rieger}} \bibnamefont{and}
  \bibinfo{author}{\bibfnamefont{N.}~\bibnamefont{Kawashima}},
  \bibinfo{journal}{Eur. Phys. J. B} \textbf{\bibinfo{volume}{9}},
  \bibinfo{pages}{233} (\bibinfo{year}{1999}).

\end{thebibliography}

\end{document}